\documentclass[conference]{IEEEtran}

\IEEEoverridecommandlockouts
\usepackage{cite}
\usepackage{amsmath,amssymb,amsfonts}
\usepackage{algorithmic}
\usepackage{listings}
\usepackage{hyperref}
\usepackage{graphicx}
\usepackage{textcomp}
\usepackage{xcolor}
\usepackage{xparse}

\definecolor{customgreen}{rgb}{0,0.6,0}
\definecolor{customgray}{rgb}{0.5,0.5,0.5}
\definecolor{custommauve}{rgb}{0.6,0,0.8}

\hypersetup{
    colorlinks=true,
    allcolors=cyan,      
    pdfpagemode=FullScreen,
}

\NewDocumentCommand{\codeword}{v}{
\texttt{\textcolor{black}{#1}}
}

\def\BibTeX{{\rm B\kern-.05em{\sc i\kern-.025em b}\kern-.08em
    T\kern-.1667em\lower.7ex\hbox{E}\kern-.125emX}}
\begin{document}

\title{Skyhook: Towards an Arrow-Native \\ Storage System}

\author{\IEEEauthorblockN{Jayjeet Chakraborty}
\IEEEauthorblockA{\textit{University of California, Santa Cruz} \\
Santa Cruz, CA, USA \\
jayjeetc@ucsc.edu} \\
\IEEEauthorblockN{Alexandru Uta}
\IEEEauthorblockA{\textit{Leiden University} \\
Leiden, Netherlands \\
a.uta@liacs.leidenuniv.nl}
\and
\IEEEauthorblockN{Ivo Jimenez}
\IEEEauthorblockA{\textit{University of California, Santa Cruz} \\
Santa Cruz, CA, USA \\
ivotron@ucsc.edu} \\
\IEEEauthorblockN{Jeff LeFevre}
\IEEEauthorblockA{\textit{University of California, Santa Cruz} \\
Santa Cruz, CA, USA \\
jlefevre@ucsc.edu} \\
\and
\IEEEauthorblockN{Sebastiaan Alvarez Rodriguez}
\IEEEauthorblockA{\textit{Leiden University} \\
Leiden, Netherlands \\
s.f.alvarez.rodriguez@umail.leidenuniv.nl} \\
\IEEEauthorblockN{Carlos Maltzahn}
\IEEEauthorblockA{\textit{University of California, Santa Cruz} \\
Santa Cruz, CA, USA \\
carlosm@ucsc.edu}
}

\maketitle

\begin{abstract}
With the ever-increasing dataset sizes, several file formats such as Parquet, ORC, and Avro have been developed to store data efficiently, save the network, and interconnect bandwidth at the price of additional CPU utilization. However, with the advent of networks supporting $25$-$100$ Gb/s and storage devices delivering $1,000,000$ reqs/sec, the CPU has become the bottleneck trying to keep up feeding data in and out of these fast devices. The result is that data access libraries executed on single clients are often CPU-bound and cannot utilize the scale-out benefits of distributed storage systems. One attractive solution to this problem is to offload data-reducing processing and filtering tasks to the storage layer. However, modifying legacy storage systems to support compute offloading is often tedious and requires an extensive understanding of the system internals. Previous approaches re-implemented functionality of data processing frameworks and access libraries for a particular storage system, a duplication of effort that might have to be repeated for different storage systems.

This paper introduces a new design paradigm that allows extending programmable object storage systems to embed existing, widely used data processing frameworks and access libraries into the storage layer with no modifications. In this approach, data processing frameworks and access libraries can evolve independently from storage systems while leveraging distributed storage systems' scale-out and availability properties. We present Skyhook, an example implementation of our design paradigm using Ceph, Apache Arrow, and Parquet. We provide a brief performance evaluation of Skyhook and discuss key results.
\end{abstract}


\begin{figure}[h]
\centering
\includegraphics[width=0.9\linewidth]{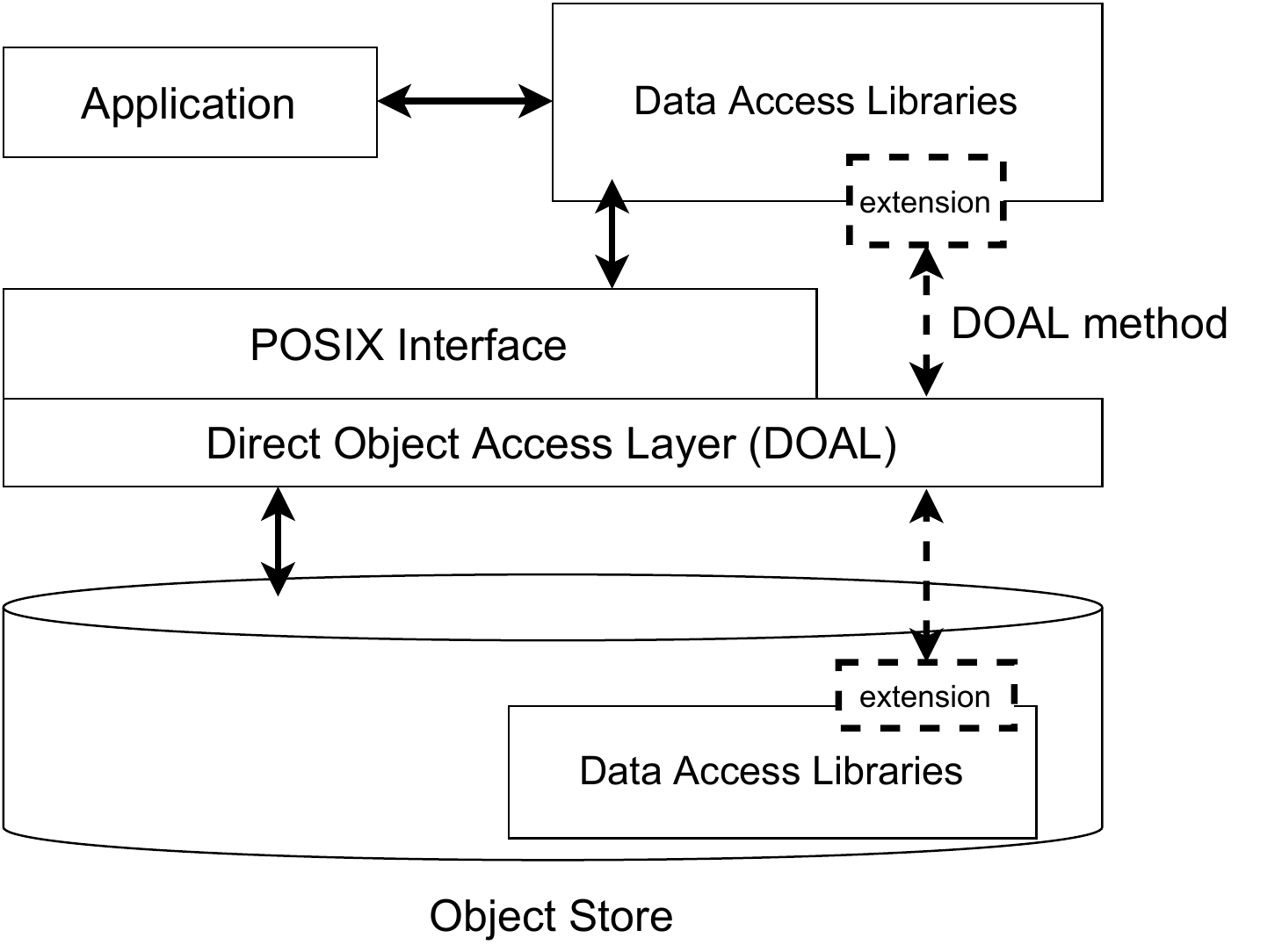}
\caption{High-level design of Skyhook.}
\label{fig:hld}
\end{figure}

\section{Introduction}
% The problem
Over the last decade, a variety of distributed data processing frameworks like Spark~\cite{zaharia2010spark} and Hadoop~\cite{white2012hadoop} have come into existence. These frameworks were built to efficiently query vast quantities of semi-structured data and get insights quickly. Unlike standard relational database management systems (RDBMS) such as MySQL~\cite{mysql}, which are optimized to manage both data storage and processing, these systems were designed to read data from a wide variety of data sources, including those in the cloud like S3~\cite{S3}. These systems depend on different file formats like Parquet~\cite{parquet}, Avro~\cite{Avro}, and ORC~\cite{Orc} for efficiently storing and accessing data. Since storage devices have been the primary bottleneck for data processing systems for a long time, the main focus of these file formats has been to store data efficiently in a binary format and reduce the amount of disk I/O required to fetch the data. However, with recent advancements in storage devices with NVMe~\cite{xu2015performance} drives and network devices with Infiniband networks~\cite{infiniband}, the bottleneck has shifted from the storage devices to the client machine's CPUs, rendering the notion of "A fast CPU and slow disk" invalid, as shown by Trivedi et al.~\cite{trivedi2018albis}. The serialized and compressed nature of these file formats makes reading them CPU-bound in systems with high-speed network and storage devices, resulting in severely reduced scalability.

% What has already been done?
An attractive solution to this problem is to offload any computation to the storage layer to achieve scalability, faster queries, and reduced network traffic. Several popular distributed data processing systems have explored this approach, e.g. IBM Netazza~\cite{singh2011introduction}, Oracle Exadata~\cite{OracleExadata}, Redshift~\cite{gupta2015amazon}, and PolarDB~\cite{cao2020polardb}. Most of these systems are built following a clean-slate approach and use specialized and costly hardware, such as Smart SSDs~\cite{do2013query} and FPGAs~\cite{fpga} for table scanning. Building systems like these requires in-depth understanding and expertise in using modern hardware for building database systems. Also, modifying existing systems like MariaDB~\cite{razzoli2014mastering}, as in the case of YourSQL~\cite{jo2016yoursql}, requires modifying code that is hardened over the years which may result in performance, security, and reliability issues. A possible way to mitigate these issues is to have programmable storage systems with low-level extension mechanisms that allow implementing application-specific data manipulation and access in their I/O path. Customizing storage systems via plugins results in minimal implementation overhead and increases the maintainability of the software. 

% What did we do?
Programmable object storage systems such as Ceph~\cite{weil2006ceph,cephwiki}, Swift~\cite{swift}, and DAOS~\cite{liang2020daos} often provide a POSIX filesystem interface for reading and writing files which are mostly built on top of object storage access libraries such as ``\codeword{librados}'' in Ceph and ``\codeword{libdaos}'' in DAOS. Being programmable, these systems provide plugin-based extension mechanisms that allow direct access and manipulation of objects within the storage layer. We leverage these features of programmable storage systems and develop a new design paradigm that allows embedding widely-used data access libraries inside the storage layer. As shown in Figure~\ref{fig:hld}, the extensions on the client and storage layers allow an application to execute data access library operations either on the client or via the direct object access layer in the storage server.

% What evaluations have we done and contributions?
We implement Skyhook, an instantiation of our design paradigm using RADOS~\cite{weil2007rados} as the programmable object storage backend, CephFS~\cite{borges2017cephfs} as the POSIX layer, Apache Arrow~\cite{Arrow} as the data access library, and Parquet as the file format. We evaluate the performance of Skyhook by scaling out the cluster measuring metrics such as query latency, CPU utilization, and network traffic. The evaluations show that Skyhook scales by offloading CPU usage for common data processing tasks to the storage layer, freeing the client for other processing tasks. Additionally, since compute units are co-located with storage nodes in Skyhook, the crash recovery and consistency semantics of the storage layer apply naturally to the query processing layer, and Skyhook queries become fault-tolerant. Since Skyhook uses Arrow heavily, we contributed our source code to the upstream Apache Arrow open-source project~\cite{skyhookinarrow}~\cite{skyhookinarrowblog}. 

In summary, our primary contributions are as follows:
\begin{itemize}
\item A design that embeds a widely-used data access library inside storage servers using a plugin infrastructure that does not require any changes to the data access library and has no impact on the storage system's resilience and failure management.

\item A design that extends a widely-used data access library with a plugin that allows offloading tasks from clients to storage servers by leveraging existing and unmodified filesystem and object storage interfaces (assuming an extensible object storage system).

\item A brief analysis of the performance gain achieved by offloading dataset scans to the storage servers. We demonstrate that offloading dataset scan operations to the storage layer results in faster queries, enhanced scalability, and reduced network traffic.
\end{itemize}

\section{Background}
\label{sec:background}

\subsection{Ceph}
Ceph is an open-source software-defined storage platform
that implements a distributed object storage layer and provides
$3$-in-$1$ interfaces for object, block, and file-level storage. One of the main properties of Ceph is that it does not have a
single point of failure because of its CRUSH~\cite{weil2006crush} map feature.
On a high-level, a CRUSH map contains object-OSD mappings, which are downloaded to the client when connecting to a Ceph cluster.
The client uses the mapping information to calculate the
location of an object in the cluster and accesses the object
directly. The object storage layer of Ceph, RADOS, is 
programmable as Ceph provides a plugin-based extension mechanism with its Object Class SDK~\cite{objectclasssdk} that allows extending the object storage layer with custom plugins written in languages such as C++ and Lua. These plugins are embedded in the form of shared libraries inside Ceph OSDs and provide the ability to access and manipulate RADOS objects on the fly within the RADOS I/O path. The Object Class SDK offers a small subset of POSIX APIs such as \codeword{read}, \codeword{write}, and \codeword{stat}. Several Ceph components, such as the RADOS Gateway (RGW) and the RADOS Block Device (RBD), use the Ceph Object Class SDK. We refer to this SDK as the Ceph CLS in the rest of the paper.

\begin{figure*}[h]
\centering
\includegraphics[width=0.9\textwidth]{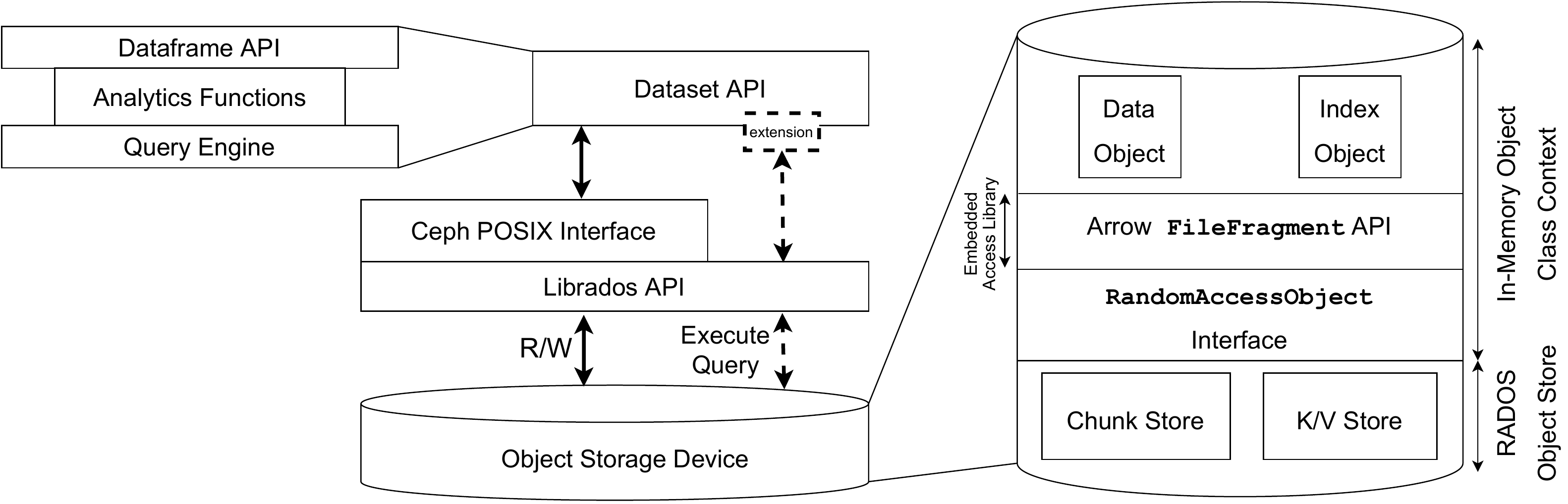}
\caption{Architecture of Skyhook.}
\label{fig:architecture}
\end{figure*}

\subsection{Apache Arrow}
 Arrow is a zero-copy columnar in-memory format for storing structured data optimized for efficient analytics operations on modern hardware. Arrow's uniform disk, memory, and wire format allow applications to overcome any serialization and deserialization overhead while transferring Arrow data between different processes and systems. Apart from being a highly-efficient in-memory format, Arrow is also a collection of several modular data processing components, collectively utilizing which allows building parts or all of a data processing system. Some of the data-processing pieces that Arrow provides are as follows: Gandiva, A LLVM-based expression compiler~\cite{gandiva}; Flight, A gRPc-based data transport interface~\cite{flight}; Dataset API, A datasets abstraction~\cite{ArrowDatasetDocs}; Arrow IPC, A binary data transfer protocol. Popular data processing systems such as Spark, Dask~\cite{Dask}, and Ray~\cite{Ray} support the Arrow in-memory format. 

\subsection{Apache Parquet}
Parquet is an open-source columnar file format for efficient and long-term storage of tabular data. Parquet supports several encoding schemes such as Dictionary and Run-Length Encoding (RLE) and compression schemes such as Snappy, LZ4, and ZSTD, which makes it highly space-efficient. One of the essential features of Parquet is its ability to push down predicates. Parquet files contain row group statistics and column offsets in their footer metadata, allowing Parquet file readers to skip over row groups and column data pages containing non-relevant data. In this way, using Parquet for data storage greatly minimizes the I/O required for reads. These characteristics of Parquet make it the de-facto choice for most data processing systems in the Hadoop ecosystem.

\section{Design and Implementation}
\label{sec:design_and_impl}
In this section, we discuss our design paradigm, the motivation behind it and provide an in-depth description of the internals of Skyhook. Additionally, we also discuss the portability of our design paradigm to other storage systems and data access libraries.

\subsection{Design Paradigm}
\label{sec:design_paradigm}
One of the most critical aspects of our design is that it allows building in-storage data processing systems with minimal implementation effort. Our design extends the client and storage layers of programmable object storage systems with widely-used data access libraries requiring absolutely no modifications. We achieve this by (1) creating a filesystem shim in the object storage layer so that data access libraries embedded in the storage layer can continue to operate on files, (2) mapping client requests-to-be-offloaded directly to objects using filesystem striping metadata, and (3) mapping files to logically self-contained fragments by using standard filesystem striping. As shown in Figure~\ref{fig:architecture}, we developed one instantiation of our design paradigm using Ceph as the storage system, Arrow as the data access library, and Parquet as the file format. We expose our implementation called Skyhook via the Arrow Dataset API by creating a new file format abstraction named the \codeword{SkyhookFileFormat}, that extends the \codeword{FileFormat} API~\cite{ArrowFileFormat} in Arrow. In the next section, we discuss the internals of Skyhook in detail.

\subsection{Skyhook: An Example Implementation}
\label{sec:implementation}

\textbf{Extending the Ceph Object Store.}
In the storage layer, using the Ceph CLS, we create a ``\codeword{scan_op}'' object class method that reads a RADOS object containing tabular data encoded in binary file formats such as Parquet and ORC, scans it using Arrow APIs by applying the query parameters received from the client and returns the scanned result in the form of an Arrow table. Reading from binary files requires a filesystem shim (a filesystem-like random access interface) to seek around in a file and read from any offset. Since the Ceph CLS does not provide a full POSIX-like API, for example, with a ``\codeword{seek}'' method, we utilize the Ceph CLS APIs to create a filesystem shim that allows seeks by keeping track of the file pointer, thereby providing a file-like view over objects. Arrow provides a \codeword{FileFragment} API that wraps a file and allows applying scan operations to it. It takes predicates and projections as inputs, applies them to a file, and returns the result in the form of an Arrow table. Since the filesystem shim allows interacting with objects as files, it seamlessly plugs into the \codeword{FileFragment} API.

\textbf{Tunneling through CephFS.} 
In Skyhook, files are written to Ceph using the CephFS POSIX interface, which stripes files over fixed-size objects stored in the RADOS layer. During the read phase, to execute the ``\codeword{scan_op}'' Ceph CLS method on a RADOS object, Skyhook must first map a file to a RADOS object. This mapping is achieved by leveraging the file striping metadata provided by the Metadata Server (MDS) of CephFS to calculate object IDs from file names. We implement a direct object access (DOA) API on the client-side that facilitates this translation and invokes the Ceph CLS method ``\codeword{scan_op}'' on the corresponding RADOS object. The DOA API takes as input a serialized flatbuffer~\cite{flatbuffers} which comprises the scan request and contains the parameters as shown in Listing~\ref{lst:scanreq}. The ``\codeword{scan_op}'' method is invoked with the serialized scan request as input. Using the DOA API, the client can interact with RADOS objects and manipulate them directly in application-specific ways while also having a filesystem view over the objects. It is almost impossible to implement reads without the DOA API because the dynamically generated result data inside the storage servers is not byte-addressable, implying a filesystem interface cannot be used.

\begin{lstlisting}[numbers=none,language=Python,label={lst:scanreq},caption={Scan request in Skyhook.}]
table ScanRequest {
  file_size: long;
  file_format: short;
  filter_expression: [ubyte];
  partition_expression: [ubyte];
  dataset_schema: [ubyte];
  projection_schema: [ubyte];
}
\end{lstlisting}

\textbf{Extending the Arrow Dataset API.} 
The Arrow framework provides a \codeword{FileFormat} API that plugs into the Dataset API~\cite{ArrowDatasetDocs} and allows scanning datasets in a unified manner. It provides APIs for scanning datasets of different file formats such as Parquet, ORC, CSV, and Feather~\cite{feather}. Since Parquet is the de-facto file format in most modern data processing systems, we use it as a baseline for our performance evaluations. We extend the \codeword{FileFormat} API in Arrow to create a \codeword{SkyhookFileFormat} API that leverages the DOA API to enable offloading fragment scan operations to Ceph OSDs. The \codeword{SkyhookFileFormat} API serializes the query parameters and passes them on to the DOA API for sending them to the storage layer. This API is generic enough to offload scan operations on every file format supported by Arrow as long as there is a Ceph CLS method to scan the particular file format in the storage layer. This API allows client applications using the Arrow Dataset API to offload file scan operations to the RADOS layer in Ceph by simply changing the file format argument in the Dataset API as shown in Listing~\ref{lst:usage}.

\begin{figure}[h]
\centering
\includegraphics[width=\linewidth]{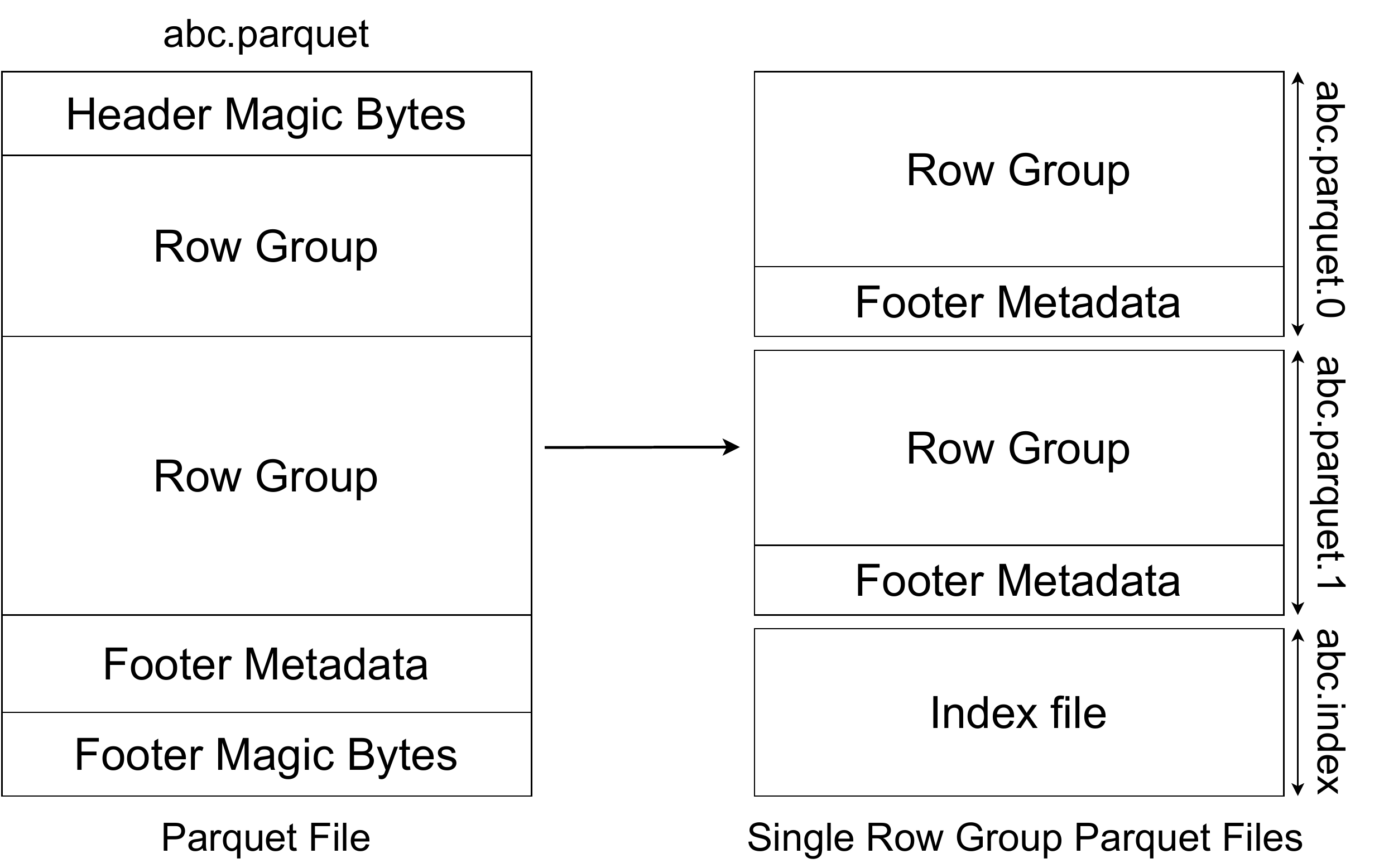}
\caption{File layout design for Skyhook.}
\label{fig:split}
\end{figure}

\textbf{Establishing a File Layout Design.}
\label{sec:file_layout_design}
Being very efficient in storing and accessing data, Parquet has become the de-facto file format for popular data processing systems like Spark and Hadoop. Since Parquet files are often multiple gigabytes in size, a standard way to store Parquet files is to store them in blocks as in HDFS~\cite{borthakur2007hadoop,HDFS}, where the typical block size is $128$\,MB~\cite{hdfsblocksize}. While writing Parquet files to HDFS, each row group is stored in a single block to prevent reading across multiple blocks when accessing a single row group. We aim to follow a similar file layout for storing Parquet files in Ceph so that every row group is self-contained within an object. In our design, a Parquet file with $R$ row groups is split into $R$ Parquet files, each containing the data and metadata from a single row group. To retain the optimizations due to the predicate pushdown capability in Parquet, the footer metadata and the schema of the parent Parquet file are also serialized and written to a separate Parquet file ending with a ".index" extension. So, for every Parquet file containing $R$ row groups, $R + 1$ small Parquet files are written, each contained within a single RADOS object. We refer to this set of $R + 1$ Parquet files as a logical Parquet file. During the dataset discovery phase, we discover only those files that end with a ".index" extension and read out the schema information from them. During the query execution phase, for every logical Parquet file, first, the footer metadata from the index file is read. Then, the IDs of the row groups that qualify for scanning are calculated based on the row group statistics present in them. Finally, the row group IDs are translated to the corresponding filenames, and then the underlying objects of these files are scanned in parallel via the DOA API. Similar to scanning Parquet files containing data from row groups, the index Parquet file is also scanned using a Ceph CLS method. The index file feature discussed here is only a design proposal and is yet to be implemented in Skyhook. Figure~\ref{fig:split} shows the file layout design.

\begin{lstlisting}[numbers=none,language=Python,label={lst:usage},caption={Reading from a Parquet dataset \textbf{with and without} Skyhook using the Arrow Dataset API.}]
# Reading from Parquet
import pyarrow.dataset as ds
format_ = "parquet"
dataset = ds.dataset(
    "/dataset", format=format_
)
dataset.to_table()

# Reading from Parquet using Skyhook
import pyarrow.dataset as ds
format_ = ds.SkyhookFileFormat(
    "parquet", "/ceph.conf"
)
dataset = ds.dataset(
    "/dataset", format=format_
)
dataset.to_table()
\end{lstlisting}

\subsection{Portability}
The implementation of Skyhook consists of several pluggable and reusable components that make it feasible to port Skyhook's design to other systems. In this section, we discuss how the design paradigm of Skyhook can be applied to other programmable object storage systems and data-access libraries by specifying the requirements in both the client and the storage layers. 

We start by discussing the portability of Skyhook's design to other programmable object storage systems. The DOA and the storage connection APIs defined in the \codeword{SkyhookFileFormat} API on the client-side needs to use the object storage system specific access library to be able to connect to the storage layer and execute object class methods in the storage nodes. On the storage-side, the filesystem shim defined in the storage plugin needs to use the storage system specific object class SDK to access the corresponding object store. Additionally, the storage system should have a POSIX interface that allows using a user-defined stripe unit while writing files. This is essential to ensure that every file is backed by a single object. 

Skyhook's design paradigm can be ported to other data access libraries. To achieve this, the data access library on the client-side should have a Dataset API to discover and instantiate a dataset abstraction over a directory of files. It should be able to map file scan calls to object class method calls using the DOA API. Additionally, a format-specific file writer is required to split large files into object-sized ones and write them in the POSIX filesystem by manipulating the stripe unit. In the storage nodes, the data access library should be embeddable in the form of shared libraries and should compile with the storage plugin. Most importantly, it needs to use the filesystem shim as a data source to read from objects as files.

\section{Evaluations}
\label{sec:evaluation}
%What is the experiment setup?
We performed experiments to compare the query latency, CPU utilization, and network bandwidth usage of accessing a dataset and filtering at the client or the storage server. The experiments were performed on CloudLab, the NSF-funded bare-metal-as-a-service infrastructure~\cite{Duplyakin+:ATC19}. For our experiments, we exclusively used machines with an $8$-core Intel Xeon D-1548 $2.0$ GHz processor (with hyperthreading enabled), $64$\,GB DRAM, a $256$\,GB NVMe drive, and a $10$\,GbE network interface. These bare-metal nodes are codenamed ``m510'' in CloudLab. We ran our experiments on Ceph clusters with $4$, $8$, and $16$ storage nodes and a single client node. Each storage node had a single Ceph OSD running on the NVMe drive. We configured the OSDs to use $8$ threads to prevent contention due to hyperthreading in the storage nodes. The number of placement groups (PGs) was increased from $128$ to $512$ while scaling out the cluster to avoid increased lock contention. A CephFS interface was created on a $3$-way replicated pool and was mounted in user mode using the ceph-fuse utility.

%What was the workload?
We used a dataset with $1.2$ billion rows and $17$ columns consisting of data from the NYC yellow taxi dataset~\cite{yellowtaxi} as our workload. The in-memory size of the dataset was found to be about $155$\,GB. In this work, we experimented with $64$\,MB files as we discovered that $64$\,MB files produced the best results for both with and without Skyhook cases. The dataset was generated by replicating a $64$\,MB Parquet file $460$ times. The Parquet files used were uncompressed. The fact that reading directly from RADOS bypasses any cache allowed us to replicate the same file without any caching-related implications on performance. Since a row group is supposed to be self-contained within a single object and the unit of parallelism in the Arrow Dataset API when using Parquet is a single row group, we used Parquet files having a single row group backed by a single RADOS object in all our experiments. Accordingly, the underlying object size was also $64$\,MB. Since the index file feature was not available in Skyhook, we used similarly configured datasets containing only single row group $64$\,MB Parquet files for both with and without Skyhook experiments. We used the Python version of the Arrow Dataset API for the experiments and used Python's ThreadPoolExecutor for launching scans in parallel, following an asynchronous I/O model. We measured latency, scalability, CPU usage, and network traffic of scanning a Parquet dataset with and without Skyhook.

\subsection{Latency}
\label{sec:eval_latency}
\begin{figure*}[h]
\centering
\includegraphics[width=\textwidth]{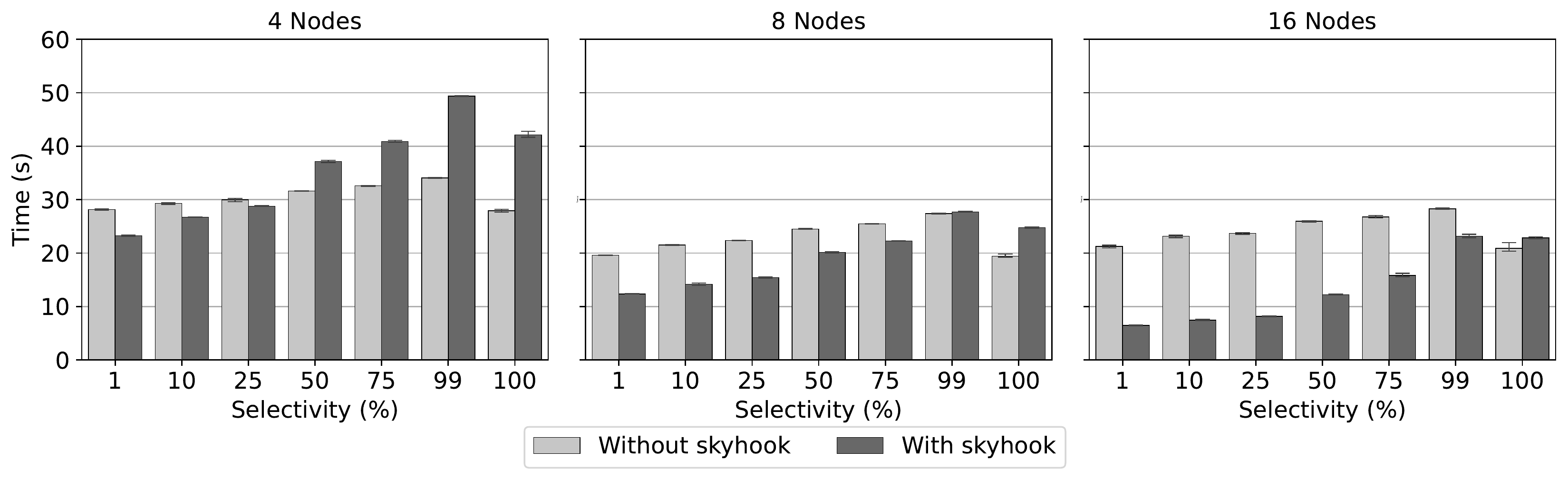}
\caption{Query latency variation on scaling out from $4$ to $8$ and $16$ storage nodes.}
\label{fig:latency}
\end{figure*}

We ran queries to select $100\%$, $99\%$, $75$\%, $50$\%, $25$\%, $10\%$, and $1\%$ of the rows from our dataset. The queries were crafted so that the rows always get selected from the beginning of the row group in every file. So, $10$\% of the rows indicates that the first $10$\% of the rows were selected from every file. In the $100\%$ selectivity case, Skyhook reads and returns all the rows without applying any filters to the dataset. The IO depth at each storage node was maintained at $2$ across all the experiments. As shown in Figure~\ref{fig:latency}, scanning with Skyhook is slower than without Skyhook in the $100\%$ scenario since there is no reduction of data sent over the network, and on top of that, Skyhook ends up sending more data as it uses LZ4-compressed Arrow IPC format on the wire which has a slightly larger byte size than the uncompressed Parquet format. On scaling out from $4$ to $8$ nodes, the performance of scans, both with and without Skyhook, improved as there were still CPU resources available to use. But, on scaling out from $8$ to $16$ nodes, the scan without Skyhook bottlenecked on the client CPU usage and stopped scaling further, whereas the scans with Skyhook, being not bottlenecked, scaled out. Skyhook maintains scalability as it can offload and distribute computation across all the storage nodes rather than staying CPU bottlenecked on the client. Also, the conversion from in-memory to wire format of Arrow requires multiple small memory copies whose overhead add up to cause significant performance degradation in Skyhook. This performance degradation is visible in the $4$ nodes scenario because the without Skyhook scan is not bottlenecked, and there is nothing to trade-off the conversion overhead of Skyhook scans as in the $8$ and $16$ nodes cases.

\subsection{CPU Usage}

\begin{figure}[h]
\centering
\includegraphics[width=\linewidth]{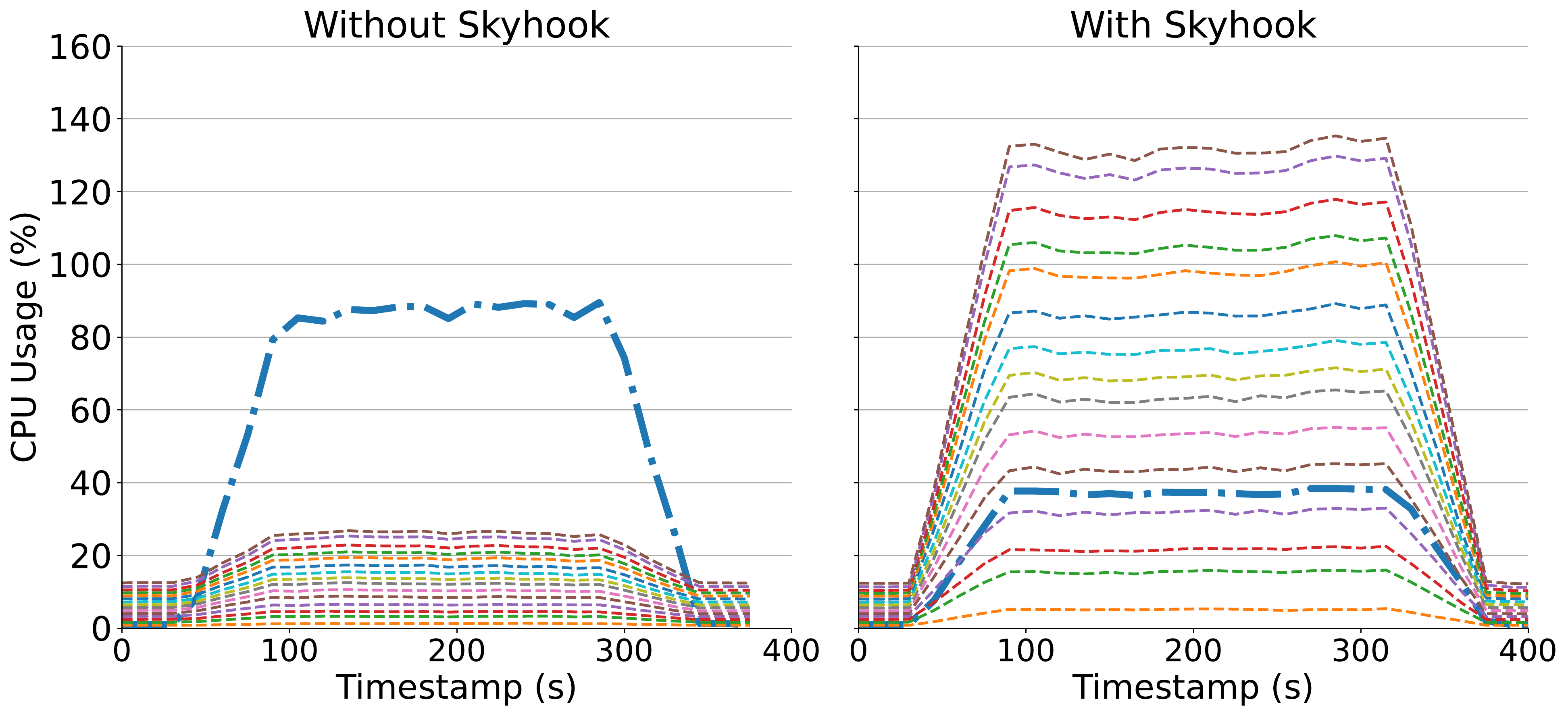}
\caption{CPU utilization of scans without (left) and with (right) Skyhook. The thick blue line shows the CPU utilization on the client, and the thin lines show the CPU utilization on the storage servers.}
\label{fig:cpu}
\end{figure}

Figure~\ref{fig:cpu} shows the CPU utilization over time by the client and the storage layers during a query execution with $100\%$ selectivity on a cluster with $16$ storage nodes. The CPU usage has been recorded using Prometheus's~\cite{prometheus} node metrics exporter. We observe that Parquet almost exhausts the client's CPU even with no filtering involved. This observation implies that the client would be unable to do any other processing work, and the query performance is bottlenecked on the client's CPU. On the other hand, we observe that with Skyhook, plenty of CPU is available on the client node, and there is high CPU utilization on the storage layer. The CPU usage on the client node can be attributed to the decompression of LZ4-compressed Arrow IPC data received as the result of scan operations. Hence, with plenty of CPU available on the client-side, more asynchronous threads can be launched to improve parallelism, or the client can take other processing tasks.

\subsection{Network Traffic}

\begin{figure}[h]
\centering
\includegraphics[width=\linewidth]{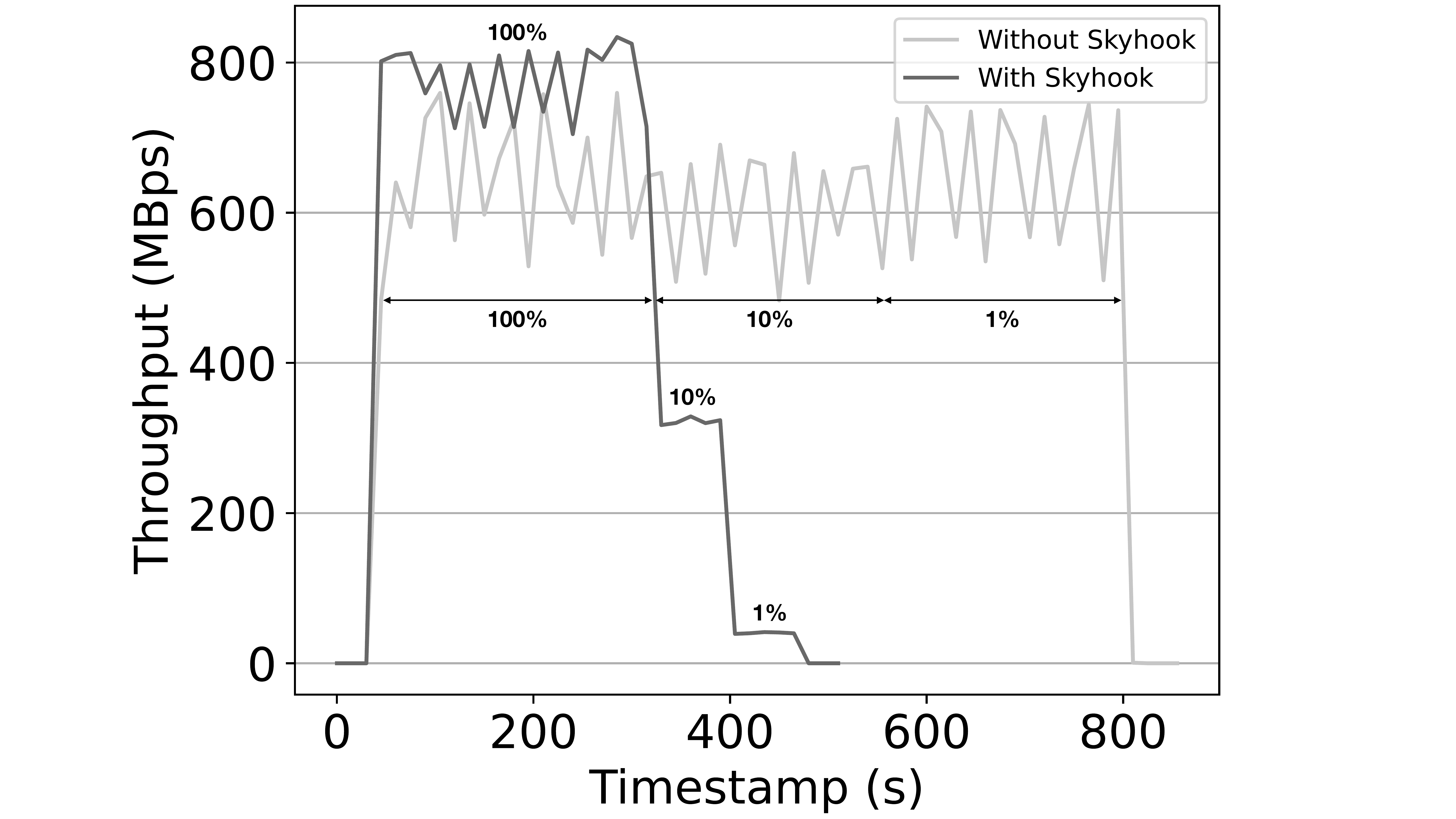}
\caption{Network bandwidth usage with and without Skyhook.}
\label{fig:network}
\end{figure}

While performing our experiments, we recorded the network traffic received on the client node. Figure~\ref{fig:network} shows the network throughput for query executions with $100$\%, $10$\%, and $1$\% selectivities on a cluster with $16$ storage nodes. On scanning with Skyhook, the network traffic keeps reducing as we move from lower ($100$\%) to higher ($1$\%) selectivities. In contrast, without Skyhook, the network traffic always remains the same, which implies significant bandwidth wastage. The numbers recorded show that using Skyhook results in significant data movement reduction. The network traffic for Skyhook is slightly higher than without Skyhook for the $100$\% scenario because the LZ4-compressed Arrow IPC format that Skyhook uses for transferring data is slightly larger in byte size than the uncompressed Parquet binary format, and moving that extra data becomes an overhead since, during $100$\% selectivity, all of the data is sent back in both cases.

\subsection{Crash Recovery}

\begin{figure}[h]
\centering
\includegraphics[width=0.8\linewidth]{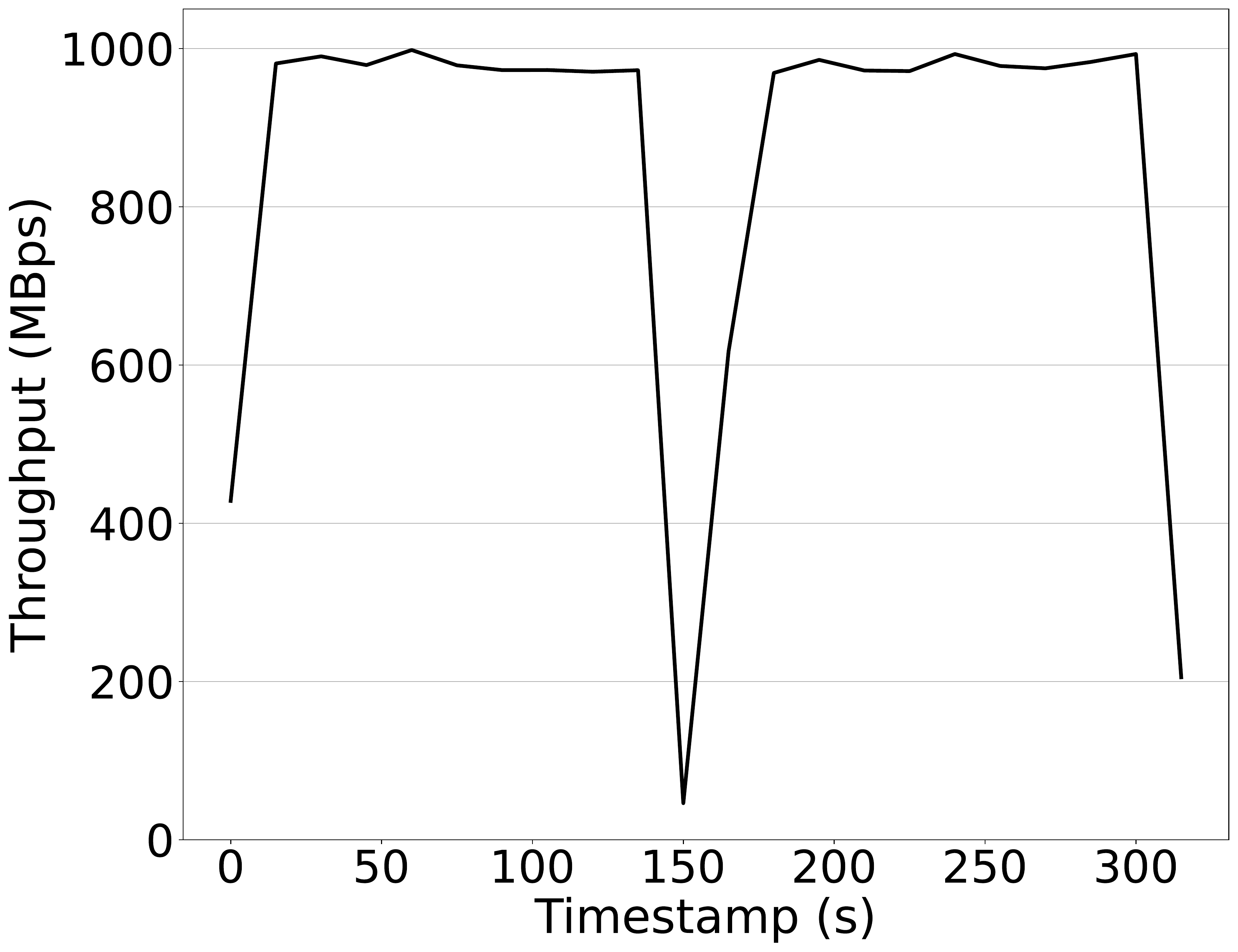}
\caption{Network throughput as perceived from the client in the event of all OSDs crashing one by one. The steep drop in throughput is where all the OSDs crashed.}
\label{fig:crash}
\end{figure}

To verify the crash recovery functionality while using Skyhook, we executed a long-running query with $100$\% row selectivity and restarted all the OSDs one-by-one in about the middle of the query execution to simulate a crash. As shown in Figure~\ref{fig:crash}, the query execution throughput resumed back to what was before the crash as the Ceph OSDs started restarting. This result shows that Skyhook queries are naturally fault-tolerant due to the co-location of query execution units with fault-tolerant storage nodes.

\section{Related Work}
Several distributed data processing systems have embraced the idea of query offloading to the storage layer for performance improvement. The paper on the Albis~\cite{trivedi2018albis} file format by Trivedi et al. explored that with high-performance storage and networks, CPUs have become the new bottleneck. Since the CPU bottleneck on the client hampers scalability, offloading CPU to the storage layer has become more important. Recently, S3 introduced S3 Select~\cite{s3select}, which allows files in either Parquet, CSV, or JSON format to be scanned inside S3 for improved query performance. Many other systems which already support reading from S3, e.g. Spark~\cite{Spark}, Redshift~\cite{gupta2015amazon}, Snowflake~\cite{Snowflake}, and PushdownDB~\cite{yu2020pushdowndb} have started taking advantage of S3 Select. But being an IaaS~\cite{iaas}, the performance of S3 Select cannot be tuned, nor can it be customized to read from file formats except the ones it supports. Systems like IBM Netezza~\cite{singh2011introduction}, PolarDB~\cite{cao2020polardb}, and Ibex~\cite{woods2014ibex} depend on sophisticated and costly hardware like FPGAs and Smart SSDs to perform table scanning inside the storage layer. These systems employ hardware-software co-design techniques to serve their specific use cases. These systems generally follow a clean-slate approach and are built from the ground up specifically tailored for query offloading. The recent work by Adams et al.~\cite{adams2021enabling} shows that near-data processing can be enabled without requiring modifications by ensuring that data elements do not cross shard boundaries and by sharing data layout information with the clients. On a high level, this approach is somewhat similar to that of Skyhook, except that the work is in its early stages, and the authors do not provide any performance evaluation of their prototype implementation. 

In our approach, we take a programmable storage system, Ceph, and extend its filesystem and object storage layers to allow offloading queries leveraging the extension mechanisms it provides. Storage systems like OpenStack Swift~\cite{swift} and DAOS~\cite{liang2020daos} also provide extension mechanisms via Storelets~\cite{storelets} and DAOS middleware~\cite{middlewaredaos} respectively. We embed Arrow libraries inside the storage layer to build the data access logic. Our approach signifies that storage systems should provide plugin mechanisms so that they can be easily extended to support ad-hoc functionality and do not need modification of legacy code or require a complete rebuild.

\section{Future Work}
As discussed in Section~\ref{sec:eval_latency}, currently, Skyhook's performance suffers from the in-memory to wire serialization overhead of Apache Arrow. We believe that employing techniques such as using RDMA for transferring data over the network can help alleviate this serialization overhead. Implementing this would require changes in the wire protocols of both Arrow and Ceph. On the use cases front, we are currently working on integrating Skyhook with in-memory SQL-based query engines such as DuckDB~\cite{duckdb} and evaluating the performance over industry-standard benchmarks such as TPC-H and TPC-DS. We also plan on assessing the performance of Skyhook with modern distributed computing systems such as Dask and Ray, as their recent addition of Dataset API support has made their integration with Skyhook seamless. Another line of future work deals with using Gandiva to accelerate the query processing performance of Arrow inside the storage layer. Finally, we plan on implementing a Flight-based communication protocol in Ceph to make the system faster and more arrow-native.

\section{Conclusion}
This paper presents a new design paradigm that allows extending the POSIX interface and the object storage layer of programmable object storage systems with plugins to allow offloading compute-heavy data processing tasks to the storage layer. We discuss how to embed different data access libraries and processing frameworks inside plugins to build a universal data processing engine that supports various file formats. We present an instantiation of this design, Skyhook, implemented using Ceph for the storage system and Arrow for the data processing layer. Currently, Skyhook supports reading Parquet files only, but we can easily add support for other file formats since we use Arrow as our data access library. We also discuss our file layout design for storing Parquet files in Ceph, similar to HDFS that allows efficient querying. We expose our implementation via a \codeword{SkyhookFileFormat} API, an extension of the Arrow \codeword{FileFormat} API, and also contributed our code to the Apache Arrow open-source project. Additionally, we present a brief performance evaluation of Skyhook and demonstrate that offloading compute-heavy query execution to the storage layer helps improve query performance by reducing data movement significantly.

\section{Acknowledgements}
This work was partially funded by the United States National Science Foundation Cooperative Agreement OAC-1836650 (IRIS-HEP), CNS-1764102, and CNS-1705021, as well as by the UC Santa Cruz Center for Research in Open Source Software (CROSS), and through the Dutch National Science Foundation NWO Veni grant VI.202.195. We would also like to thank the members of the Apache Arrow community for their extensive reviews on our pull request and to everyone who has contributed code to the project. Finally, we would like to thank Dr. Julian Kunkel for shepherding this paper and the anonymous reviewers for their valuable feedback.

\bibliographystyle{IEEEtran}
\bibliography{conference_101719}

\begin{thebibliography}{10}
\providecommand{\url}[1]{#1}
\csname url@samestyle\endcsname
\providecommand{\newblock}{\relax}
\providecommand{\bibinfo}[2]{#2}
\providecommand{\BIBentrySTDinterwordspacing}{\spaceskip=0pt\relax}
\providecommand{\BIBentryALTinterwordstretchfactor}{4}
\providecommand{\BIBentryALTinterwordspacing}{\spaceskip=\fontdimen2\font plus
\BIBentryALTinterwordstretchfactor\fontdimen3\font minus
  \fontdimen4\font\relax}
\providecommand{\BIBforeignlanguage}[2]{{%
\expandafter\ifx\csname l@#1\endcsname\relax
\typeout{** WARNING: IEEEtran.bst: No hyphenation pattern has been}%
\typeout{** loaded for the language `#1'. Using the pattern for}%
\typeout{** the default language instead.}%
\else
\language=\csname l@#1\endcsname
\fi
#2}}
\providecommand{\BIBdecl}{\relax}
\BIBdecl

\bibitem{zaharia2010spark}
M.~Zaharia, M.~Chowdhury, M.~J. Franklin, S.~Shenker, I.~Stoica \emph{et~al.},
  ``Spark: Cluster computing with working sets.'' \emph{HotCloud}, vol.~10, no.
  10-10, p.~95, 2010.

\bibitem{white2012hadoop}
T.~White, \emph{Hadoop: The definitive guide}.\hskip 1em plus 0.5em minus
  0.4em\relax " O'Reilly Media, Inc.", 2012.

\bibitem{mysql}
``Mysql,'' \url{https://www.mysql.com/}.

\bibitem{S3}
Amazon, ``Amazon s3,'' \url{https://aws.amazon.com/s3/}, Amazon, accessed:
  2020-11-16.

\bibitem{parquet}
``Apache parquet,'' \url{https://parquet.apache.org/}, Apache Software
  Foundation.

\bibitem{Avro}
Apache, ``Apache avro,'' \url{https://avro.apache.org/}, Apache Software
  Foundation.

\bibitem{Orc}
``Apache orc: High-performance columnar storage for hadoop,'' Apache Software
  Foundation, 2018.

\bibitem{xu2015performance}
Q.~Xu, H.~Siyamwala, M.~Ghosh, T.~Suri, M.~Awasthi, Z.~Guz, A.~Shayesteh, and
  V.~Balakrishnan, ``Performance analysis of nvme ssds and their implication on
  real world databases,'' in \emph{Proceedings of the 8th ACM International
  Systems and Storage Conference}, 2015, pp. 1--11.

\bibitem{infiniband}
G.~F. Pfister, ``An introduction to the infiniband architecture,'' \emph{High
  performance mass storage and parallel I/O}, vol.~42, no. 617-632, p.~10,
  2001.

\bibitem{trivedi2018albis}
A.~Trivedi, P.~Stuedi, J.~Pfefferle, A.~Schuepbach, and B.~Metzler, ``Albis:
  High-performance file format for big data systems,'' in \emph{2018
  $\{$USENIX$\}$ Annual Technical Conference ($\{$USENIX$\}$$\{$ATC$\}$ 18)},
  2018, pp. 615--630.

\bibitem{singh2011introduction}
M.~Singh and B.~Leonhardi, ``Introduction to the ibm netezza warehouse
  appliance,'' in \emph{Proceedings of the 2011 Conference of the Center for
  Advanced Studies on Collaborative Research}, 2011, pp. 385--386.

\bibitem{OracleExadata}
Oracle, ``Oracle exadata,''
  \url{https://www.oracle.com/engineered-systems/exadata/ }, Oracle.

\bibitem{gupta2015amazon}
A.~Gupta, D.~Agarwal, D.~Tan, J.~Kulesza, R.~Pathak, S.~Stefani, and
  V.~Srinivasan, ``Amazon redshift and the case for simpler data warehouses,''
  in \emph{Proceedings of the 2015 ACM SIGMOD international conference on
  management of data}, 2015, pp. 1917--1923.

\bibitem{cao2020polardb}
W.~Cao, Y.~Liu, Z.~Cheng, N.~Zheng, W.~Li, W.~Wu, L.~Ouyang, P.~Wang, Y.~Wang,
  R.~Kuan \emph{et~al.}, ``$\{$POLARDB$\}$ meets computational storage:
  Efficiently support analytical workloads in cloud-native relational
  database,'' in \emph{18th $\{$USENIX$\}$ Conference on File and Storage
  Technologies ($\{$FAST$\}$ 20)}, 2020, pp. 29--41.

\bibitem{do2013query}
J.~Do, Y.-S. Kee, J.~M. Patel, C.~Park, K.~Park, and D.~J. DeWitt, ``Query
  processing on smart ssds: Opportunities and challenges,'' in
  \emph{Proceedings of the 2013 ACM SIGMOD International Conference on
  Management of Data}, 2013, pp. 1221--1230.

\bibitem{fpga}
Z.~A.~O. Nasri~Sulaiman, M.~Marhaban, and M.~Hamidon, ``Design and
  implementation of fpga-based systems-a review,'' \emph{Australian Journal of
  Basic and Applied Sciences}, vol.~3, no.~4, pp. 3575--3596, 2009.

\bibitem{razzoli2014mastering}
F.~Razzoli, \emph{Mastering MariaDB}.\hskip 1em plus 0.5em minus 0.4em\relax
  Packt Publishing Ltd, 2014.

\bibitem{jo2016yoursql}
I.~Jo, D.-H. Bae, A.~S. Yoon, J.-U. Kang, S.~Cho, D.~D. Lee, and J.~Jeong,
  ``Yoursql: a high-performance database system leveraging in-storage
  computing,'' \emph{Proceedings of the VLDB Endowment}, vol.~9, no.~12, pp.
  924--935, 2016.

\bibitem{weil2006ceph}
S.~A. Weil, S.~A. Brandt, E.~L. Miller, D.~D. Long, and C.~Maltzahn, ``Ceph: A
  scalable, high-performance distributed file system,'' in \emph{Proceedings of
  the 7th symposium on Operating systems design and implementation}, 2006, pp.
  307--320.

\bibitem{cephwiki}
Wikipedia, ``Ceph (software),'' \url{https://bit.ly/3mkapHU}.

\bibitem{swift}
OpenStack, ``Openstack swift storlet engine overview,''
  \url{https://docs.openstack.org/storlets/latest/}, OpenStack.

\bibitem{liang2020daos}
Z.~Liang, J.~Lombardi, M.~Chaarawi, and M.~Hennecke, ``Daos: A scale-out high
  performance storage stack for storage class memory,'' in \emph{Asian
  Conference on Supercomputing Frontiers}.\hskip 1em plus 0.5em minus
  0.4em\relax Springer, 2020, pp. 40--54.

\bibitem{weil2007rados}
S.~A. Weil, A.~W. Leung, S.~A. Brandt, and C.~Maltzahn, ``Rados: a scalable,
  reliable storage service for petabyte-scale storage clusters,'' in
  \emph{Proceedings of the 2nd international workshop on Petascale data
  storage: held in conjunction with Supercomputing'07}, 2007, pp. 35--44.

\bibitem{borges2017cephfs}
G.~Borges, S.~Crosby, and L.~Boland, ``Cephfs: a new generation storage
  platform for australian high energy physics,'' in \emph{Journal of Physics:
  Conference Series}, vol. 898, no.~6.\hskip 1em plus 0.5em minus 0.4em\relax
  IOP Publishing, 2017, p. 062015.

\bibitem{Arrow}
A.~D. Team, ``Apache arrow,'' \url{https://arrow.apache.org}, 10 2018.

\bibitem{skyhookinarrow}
``Skyhook in apache arrow,'' \url{https://git.io/JDuxJ}, Apache Software
  Foundation.

\bibitem{skyhookinarrowblog}
``Skyhook: Bringing computation to storage with apache arrow,''
  \url{https://arrow.apache.org/blog/2022/01/31/skyhook-bringing-computation-to-storage-with-apache-arrow/},
  Apache Arrow.

\bibitem{weil2006crush}
S.~A. Weil, S.~A. Brandt, E.~L. Miller, and C.~Maltzahn, ``Crush: Controlled,
  scalable, decentralized placement of replicated data,'' in \emph{SC'06:
  Proceedings of the 2006 ACM/IEEE Conference on Supercomputing}.\hskip 1em
  plus 0.5em minus 0.4em\relax IEEE, 2006, pp. 31--31.

\bibitem{objectclasssdk}
``Ceph object class sdk,''
  \url{https://docs.ceph.com/en/latest/rados/api/objclass-sdk/\#sdk-for-ceph-object-classes
  }, Ceph.

\bibitem{gandiva}
``Gandiva: A llvm-based analytical expression compiler for apache arrow,''
  \url{https://arrow.apache.org/blog/2018/12/05/gandiva-donation/}, Apache
  Arrow, 2018.

\bibitem{flight}
``Introducing apache arrow flight: A framework for fast data transport,''
  \url{https://arrow.apache.org/blog/2019/10/13/introducing-arrow-flight/},
  Apache Arrow, 2019.

\bibitem{ArrowDatasetDocs}
``Arrow dataset api,'' \url{https://arrow.apache.org/docs/python/dataset.html},
  Apache Software Foundation.

\bibitem{Dask}
M.~Rocklin, ``Dask: Parallel computation with blocked algorithms and task
  scheduling,'' in \emph{Proceedings of the 14th python in science conference},
  vol. 126.\hskip 1em plus 0.5em minus 0.4em\relax Citeseer, 2015.

\bibitem{Ray}
P.~Moritz, R.~Nishihara, S.~Wang, A.~Tumanov, R.~Liaw, E.~Liang, M.~Elibol,
  Z.~Yang, W.~Paul, M.~I. Jordan, and I.~Stoica, ``Ray: A distributed framework
  for emerging ai applications,'' in \emph{Proceedings of the 13th USENIX
  Conference on Operating Systems Design and Implementation}, ser.
  OSDI'18.\hskip 1em plus 0.5em minus 0.4em\relax USA: USENIX Association,
  2018, p. 561–577.

\bibitem{ArrowFileFormat}
``Arrow file format,''
  \url{https://arrow.apache.org/docs/cpp/api/dataset.html#file-formats}, Apache
  Software Foundation.

\bibitem{flatbuffers}
Google, ``Google flatbuffers,'' \url{https://google.github.io/flatbuffers/},
  Google.

\bibitem{feather}
H.~Wickham, ``Feather: A fast on-disk format for data frames for r and python,
  powered by apache arrow,'' \url{https://www.rstudio.com/blog/feather/}.

\bibitem{borthakur2007hadoop}
D.~Borthakur, ``The hadoop distributed file system: Architecture and design,''
  \emph{Hadoop Project Website}, vol.~11, no. 2007, p.~21, 2007.

\bibitem{HDFS}
K.~Shvachko, H.~Kuang, S.~Radia, and R.~Chansler, ``The hadoop distributed file
  system,'' in \emph{2010 IEEE 26th symposium on mass storage systems and
  technologies (MSST)}.\hskip 1em plus 0.5em minus 0.4em\relax Ieee, 2010, pp.
  1--10.

\bibitem{hdfsblocksize}
``Hdfs block size,''
  \url{http://hadoop.apache.org/docs/r3.0.0/hadoop-project-dist/hadoop-hdfs/HdfsDesign.html},
  Apache Software Foundation.

\bibitem{Duplyakin+:ATC19}
\BIBentryALTinterwordspacing
D.~Duplyakin, R.~Ricci, A.~Maricq, G.~Wong, J.~Duerig, E.~Eide, L.~Stoller,
  M.~Hibler, D.~Johnson, K.~Webb, A.~Akella, K.~Wang, G.~Ricart, L.~Landweber,
  C.~Elliott, M.~Zink, E.~Cecchet, S.~Kar, and P.~Mishra, ``The design and
  operation of {CloudLab},'' in \emph{Proceedings of the {USENIX} Annual
  Technical Conference (ATC)}, Jul. 2019, pp. 1--14. [Online]. Available:
  \url{https://www.flux.utah.edu/paper/duplyakin-atc19}
\BIBentrySTDinterwordspacing

\bibitem{yellowtaxi}
``Nyc yellow taxi trip record data,''
  \url{https://www1.nyc.gov/site/tlc/about/tlc-trip-record-data.page }.

\bibitem{prometheus}
``Prometheus,'' \url{https://prometheus.io/}.

\bibitem{s3select}
R.~Hunt, ``S3 select and glacier select – retrieving subsets of objects,''
  \url{https://aws.amazon.com/blogs/aws/s3-glacier-select/ }, 11 2017.

\bibitem{Spark}
M.~Zaharia, M.~Chowdhury, M.~J. Franklin, S.~Shenker, I.~Stoica \emph{et~al.},
  ``Spark: Cluster computing with working sets.'' \emph{HotCloud}, vol.~10, no.
  10-10, p.~95, 2010.

\bibitem{Snowflake}
B.~Dageville, T.~Cruanes, M.~Zukowski, V.~Antonov, A.~Avanes, J.~Bock,
  J.~Claybaugh, D.~Engovatov, M.~Hentschel, J.~Huang \emph{et~al.}, ``The
  snowflake elastic data warehouse,'' in \emph{Proceedings of the 2016
  International Conference on Management of Data}, 2016, pp. 215--226.

\bibitem{yu2020pushdowndb}
X.~Yu, M.~Youill, M.~Woicik, A.~Ghanem, M.~Serafini, A.~Aboulnaga, and
  M.~Stonebraker, ``Pushdowndb: Accelerating a dbms using s3 computation,'' in
  \emph{2020 IEEE 36th International Conference on Data Engineering
  (ICDE)}.\hskip 1em plus 0.5em minus 0.4em\relax IEEE, 2020, pp. 1802--1805.

\bibitem{iaas}
R.~Moreno-Vozmediano, R.~S. Montero, and I.~M. Llorente, ``Iaas cloud
  architecture: From virtualized datacenters to federated cloud
  infrastructures,'' \emph{Computer}, vol.~45, no.~12, pp. 65--72, 2012.

\bibitem{woods2014ibex}
L.~Woods, Z.~Istv{\'a}n, and G.~Alonso, ``Ibex: An intelligent storage engine
  with support for advanced sql offloading,'' \emph{Proceedings of the VLDB
  Endowment}, vol.~7, no.~11, pp. 963--974, 2014.

\bibitem{adams2021enabling}
I.~F. Adams, N.~Agrawal, and M.~P. Mesnier, ``Enabling near-data processing in
  distributed object storage systems,'' in \emph{Proceedings of the 13th ACM
  Workshop on Hot Topics in Storage and File Systems}, 2021, pp. 28--34.

\bibitem{storelets}
OpenStack, ``Storlet's documentation,''
  \url{https://docs.openstack.org/storlets/latest/}, OpenStack.

\bibitem{middlewaredaos}
Intel, ``Daos client apis, tools and i/o middleware,''
  \url{https://github.com/daos-stack/daos/blob/master/src/README.md\#client-apis-tools-and-io-middleware},
  Intel.

\bibitem{duckdb}
M.~Raasveldt and H.~M{\"u}hleisen, ``Data management for data science-towards
  embedded analytics.'' in \emph{CIDR}, 2020.

\end{thebibliography}

\end{document}